\documentclass[aps,showpacs,preprint,pre,superscriptaddress]{revtex4}
\usepackage{graphicx,epsfig}
\input amssym.tex

\begin{document}

\title{Statistical Analysis of Composite Spectra}
\author{A.Y. Abul-Magd}
\affiliation{Faculty of Science, Zagazig University, Zagazig, Egypt}
\author{H. L. Harney}
\affiliation{Max--Planck--Institut f\"{u}r Kernphysik, Heidelberg}
\author{M.H. Simbel}
\affiliation{Faculty of Science, Zagazig University, Zagazig, Egypt}
\author{H.A. Weidenm\"{u}ller}
\affiliation{Max--Planck--Institut f\"{u}r Kernphysik, Heidelberg}
\date{\today}

\begin{abstract}
We consider nearest--neighbor spacing distributions of composite
ensembles of levels. These are obtained by combining independently
unfolded sequences of levels containing only few levels each. Two
problems arise in the spectral analysis of such data. One problem lies
in fitting the nearest--neighbor spacing distribution to the histogram
of level spacings obtained from the data. We show that the method of
Bayesian inference is superior to this procedure. The second problem
occurs when one unfolds such short sequences. We show that the
unfolding procedure generically leads to an overestimate of the
chaoticity parameter. This trend is absent in the presence of
long--range level correlations. Thus, composite ensembles of levels
from a system with long--range spectral stiffness yield reliable
information about the chaotic behavior of the system. 
\end{abstract}
\maketitle

\section{Introduction}
\label{intro} 
The statistical analysis of spectra aims at a comparison of the
spectral fluctuation properties of a given physical system with 
theoretical predictions like those of random--matrix theory (RMT),
those for integrable systems, or interpolations between these two
limiting cases.

Specific problems arise whenever the spectra under consideration
involve a relatively small number of levels. This is the situation in
the analysis of spectra of nuclei in the ground--state domain~\cite
{aw,egidy,mitchell,raman,shriner,as,garrett,enders,shriner1}, of
atomic spectra~\cite {rosenzweig,camarda,flambaum}, and of molecular
spectra~\cite {haller,zimmermann,leitner}. Here, one usually deals
with sequences of levels of the same spin and parity containing only 5
or 10 levels. Several or many such sequences are then combined to
obtain an ensemble of statistically relevant size. The sequences
forming the ensemble may involve levels of different spin--parity
and/or levels from different nuclei. The resulting data set is 
typically analysed with regard to the nearest--neighbor spacing (NNS)
distribution only. In view of the shortness of the individual
sequences, correlations between spacings of levels are not
investigated.

In the present paper, we address two problems which arise in the
analysis of such data. First, we ask whether a fit to a histogram of
the NNS distribution is the optimal way to analyze the data. We
compare this method with the method of Bayesian inference which has
been successfully used to analyze the statistical properties of
coupled microwave resonators~\cite {barbosa,dembo}. Second, for a
reliable analysis, one has to ''unfold'' the individual sequences.
This yields a new data set with mean level spacing unity. Then, one
combines these level sequences to form a larger ensemble of spacings
suitable for the statistical analysis. How big is the statistical
error due to this unfolding procedure? We answer this question for two
extreme cases where the spacings are taken from a spectrum without
(with) the long--range rigidity typical for chaotic systems,
respectively.

In Section~\ref{nns}, we give a brief summary of the NNS distribution
and of spectral analyses using it. In Section~\ref{baye}, we give a
short account of Bayesian inference tailored to the problems just
mentioned. In Section~\ref{anal}, we address the above--mentioned two
problems. Section~\ref{summ} contains a summary and our conclusions.

\section{The NNS distribution}
\label{nns}

The canonical ensembles of random--matrix theory (RMT)~\cite{mehta}
are classified according to their symmetries. Here, we focus attention
on systems which are invariant under time reversal and under space
rotations. Such systems are represented by the Gaussian orthogonal
ensemble (GOE) of random matrices. The NNS distribution of levels of
the GOE is well approximated by Wigner's surmise~\cite{wigner}
\begin{equation}
p_{{\rm {W}}}(s) = \frac{\pi}{2} s\exp\left(-\frac{\pi}{4} s^{2}
\right)\, .
\label{0}
\end{equation}
Here, $s$ is the spacing of neighboring levels in units of the mean
level spacing.

RMT was introduced originally to describe the spectral fluctuation
properties of complex quantum systems. Later, it has been conjectured~
\cite{bohigas} that RMT also applies to quantum systems whose
classical counterpart is chaotic. This conjecture has enormously
widened the range of applications of RMT~\cite{guhr,alhasid} and has
led to a juxtaposition of RMT and of the theoretical description of
quantum systems which are integrable in the classical limit. The
latter possess a NNS distribution which is generically given by the
Poisson distribution, 
\begin{equation}
p_{{\rm {P}}}(s)=\exp \left( -s\right)\, .  \label{p}
\end{equation}

There also exist intermediate situations. Examples are (i) mixed
systems where the motion in some parts of classical phase space is
regular and in other parts, chaotic, see Refs.~\cite
{guhr,izrailev,abul,prosen} and references therein; (ii)
pseudointegrable systems which possess singularities and are
integrable in the absence of these singularities, see e.g. Refs.~
\cite{richens,biswas,date}; (iii) fully chaotic systems where a
conserved symmetry is dynamically broken, or is ignored.

Here we focus attention on case (iii). The Hamiltonian for a system
with strictly conserved symmetry is block--diagonal. Each block is
characterized by a quantum number (or a set of quantum numbers) of the
symmetry under consideration and may be separately considered as a
member of a GOE. Symmetry breaking is modelled by introducing
off--diagonal blocks that couple diagonal blocks with different
quantum numbers. The resulting spectrum differs from GOE predictions.
Such modelling has been useful in the following cases. (i) Isospin
mixing in nuclear spectra and reactions~\cite{harney}. (ii) Isospin
mixing in the low--lying states of $^{26}$Al~\cite{shriner}. (iii) The
gradual breaking of a point--group symmetry (which is statistically
fully equivalent to the breaking of a quantum number like isospin) in
an experiment with monocrystalline quartz blocks~\cite{ellegaard}.
(iv) The electromagnetic coupling of the resonances in two
superconducting microwave resonators~\cite{alt}. The statistical
analyses relating to some of these cases can be found in Refs.~
\cite{aw,egidy,mitchell,raman,shriner,as,garrett,enders,shriner1}
and~\cite{alaa,barbosa,dembo}. Ignoring a conserved symmetry leads to
a superposition of several GOE spectra and to spectral
fluctuation properties which are similar to those of the cases
considered above. Very strong mixing of states possessing
different symmetries leads to a GOE distribution, and the
superposition of many GOE spectra
leads to a NNS distribution which is Poissonian.

Thus, there is a variety of physical situations that give rise to
level fluctuations which are intermediate between the GOE and the
Poisson cases. We attempt to describe the NNS distribution in all
these intermediate situations by a one--parameter family of functions
interpolating between expressions~(\ref{0}) and (\ref{p}). This family
is defined in Section~\ref{prop}. We are aware of the fact that our
procedure cannot be exact. Arguments will be given to justify its use
but it remains an approximation.

\section{Bayesian analysis of NNS distributions}
\label{baye}

The Bayesian analysis of the NNS distribution proceeds in three steps.
(i) We propose a probability distribution $p(s,f)$ for the observed
spacings $s$ of nearest neighbors. This function depends
parametrically upon the parameter $f$ which measures the deviation
from GOE statistics. It is our aim to determine $f$ from the
data. (ii) We determine the posterior distribution $P(f|s)$ for the
parameter $f$. (iii) We deduce the optimum value of $f$ together with
its statistical error. The three steps are outlined in the three
subsections which follow.

\subsection{Proposed NNS distribution}
\label{prop} 

To construct $p(s,f)$, we consider a spectrum $S$ containing levels of
the same spin and parity. (In practice, we usually deal with a set of
spectra but consider only a single one in the present Section. The
generalisation to a set of spectra is considered in Section~\ref{4.2}).
The levels in $S$ may, however, differ in other conserved quantum
numbers which are either unknown or ignored. The spectrum $S$ can then
be broken up into $m$ sub--spectra $S_{j}$ of independent sequences of
levels, with $j = 1, \ldots, m$. The fractional level number of
$S_{j}$ is denoted by $f_{j}$ where $0 < f_{j} \leq 1$ and
$\sum_{j=1}^{m} f_{j} = 1 \, .$ Let $p_{j}(s),\, j=1\dots m,$ denote
the NNS distribution for the sub--spectrum $S_{j}$. We assume that each
of the distributions $p_{j}(s)$ is given by the GOE and has unit mean
level spacing. To an excellent approximation, the $p_{j}$'s are then
given by Wigner's surmise~(1). 

The construction of the NNS distribution $p(s, f_{1}, \ldots, f_{m -
  1})$ for the superposition spectrum $S$ (with unit mean level
spacing) from the $p_{j}(s)$'s was explicitly carried out by
Rosenzweig and Porter~\cite{rosenzweig}. Their construction is not
useful in the present context because in practice, we do not know the
number $m$ of sub--spectra, nor is it possible to determine all the
parameters $f_{j}$ from the data. To overcome this difficulty, we use
an approximation scheme first proposed in Ref.~\cite{as1} which leads
to an approximate NNS distribution for $S\, ,$ viz. 
\begin{eqnarray}
p(s,f) & = & \left[ 1-f+f \left( 0.7+0.3f\right) \frac{\pi s}{2}\right]
\nonumber \\
&& \times \exp \left \{ - \left(1 - f \right) s - f \left ( 0.7 + 0.3
f \right) \frac{\pi s^{2} }{4} \right\} \ .
\label{10}
\end{eqnarray}
This function depends on only a single parameter, the mean fractional
level number $f = \sum_{j=1}^{m} f_{j}^{2}$ for the superimposed
subspectra. This quantity will eventually be used as a fit parameter.
The derivation of Eq.~(\ref{10}) and the definition of $f$ are
discussed in Appendix~\ref{app1}.

For a large number $m$ of sub--spectra, $f$ is of the order of $1/m$
and, thus, small. In this limit, $p(s,f)$ approaches $p(s,0) = p_{{\rm
    P}}(s)$ given by Eq.~(2). This expresses the well--known fact that
the superposition of many GOE level sequences produces a Poissonian
sequence. On the other hand, when $f\rightarrow 1$, $p(s,f)$
approaches the NNS of the GOE. This is why we refer to $f$ as to the
chaoticity parameter. We use $p(s, f)$ as defined above for the
analysis of the data.

Our model for $p(s, f)$ has been constructed with case (iii) of
Section~\ref{nns} in mind, even when the symmetries are only weakly
broken. In that case, the distribution~(\ref{10}) is not accurate for
very small spacings because $p(s, f)$ differs from zero at $s = 0$
while the symmetry--breaking interaction lifts all degeneracies.
However, this defect should not affect the spacing distribution beyond
the domain of very small spacings. The magnitude of this domain
depends on the ratio of the strength of the symmetry--breaking
interaction to the mean level spacing.

For the other cases of intermediate situations mentioned in
Section~\ref{nns}, our model may not be the best choice. The
experimental NNS distributions of mixed systems (case (i)) are
frequently analyzed using Brody's interpolation formula~\cite{brody},
$p(s)=a(\gamma +1)s^{\gamma }\exp \left( -as^{\gamma +1}\right) $,
where $\gamma $ is a fit parameter and $a = \left[ \Gamma \left(
\frac{\gamma +2}{\gamma +1}\right) \right] ^{\gamma +1}$, although
Eq.~(\ref{10}) often also provides a reasonable representation,
see Refs.~\cite{abul,as,as1}. Case (ii) (pseudointegrable systems)
cannot be described in the framework of the present model. These
systems are successfuly described by the so--called semi--Poisson
statistics~\cite {bogomolny} where the NNS distribution is $p(s) = 2s
\exp (-4s)$.

\subsection{Posterior Distribution}
\label{join} 

We turn to the second step of Bayesian inference, the calculation of
the posterior distribution $P(f|{\bf s})$ for $f$ given a set ${\bf s}
= (s_{1},s_{2},...,s_{N})$ of $N$ spacings $s_j$ with $j = 1, \ldots,
N$. We take the experimental $s_{j}$ to be statistically independent.
This assumption does not apply in general. Indeed, the GOE produces
significant correlations between subsequent spacings. However, the
correlations become small for spacings that are separated by a few
levels. Under this assumption, the conditional probability
distribution $p\left({\bf s} \left|f\right.\right)$ of the set of
spacings ${\bf s}$ is given by
\begin{equation}
p\left({\bf s}\left|f\right.\right) =\prod_{i=1}^{N}p(s_{i},f)\, ,
\label{12}
\end{equation}
with $p(s_{i},f)$ given by Eq.~(3). Bayes' theorem then provides the
posterior distribution 
\begin{equation}
P(f|{\bf s})=\frac{p({\bf s}|f)\mu (f)}{M({\bf s})}
\label{13}
\end{equation}
of the parameter $f$ given the events ${\bf s}$. Here, $\mu (f)$ is
the prior distribution and 
\begin{equation}
M({\bf s})=\int_{0}^{1} p\left({\bf
  s}\left|f\right.\right)\mu\left(f\right) \, {\rm d}f
\label{14}
\end{equation}
is the normalization.

To find the prior distribution, we use Jeffreys' rule~\cite{jeffreys}
discussed in Chap. 9 of Ref.~\cite{harneyb},
\begin{equation}
\mu (f)\propto \biggl| \int p\left({\bf s}\left|f\right. \right) \
\left[ \partial\ln p\left({\bf s}\left|f\right. \right)/\partial f
\right]^{2} \ {\rm d}{\bf s}\ \biggr|^{1/2}\, .
\label{15}
\end{equation}
This formula --- suggested 60 years ago --- has been justified later
through a combination of arguments based on symmetry and differential
geometry~\cite{kass,harneyb}. The prior distribution may be
intuitively interpreted as the distribution ascribed to $f$ in the
absence of any observed $s\, .$ However, this interpretation requires
additional arguments when $\mu$ is an improper distribution. A
mathematically more satisfactory interpretation of $\mu$ considers
$\mu$ as the measure in the space of $f\, .$ This interpretation
justifies Jeffreys' rule as that rule yields the invariant measure of
a suitable symmetry group.

Details on the posterior distribution can be found in Appendix
\ref{app2}.

\subsection{Best--fit value for $f$}
\label{best} 

The third and last step of the Bayesian analysis consists in
determining the best--fit value and the associated error of the
chaoticity parameter $f$ for each NNS distribution. When
$P\left(f\left|{\bf s}\right.\right)$ is not Gaussian, the best--fit
value of $f$ cannot be taken as the most probable value. Rather we
take the best--fit value to be the mean value $\overline{f}$ and
measure the error by the standard deviation $\sigma $ of the posterior
distribution~(\ref{22}), i.e. 
\begin{equation}
\overline{f}=\int_{0}^{1} fP\left(f\left|{\bf s}\right.\right){\rm d}f
{\rm \ \ \ and \ \ } \sigma^2  = \int_{0}^{1}
\left(f-\overline{f}\right)^{2} P\left(f\left|{\bf
    s}\right.\right){\rm d}f\, .
\label{23}
\end{equation}
The second equation~(\ref{23}) does not provide the optimal definition
of an error interval, see Chap. 3 of~\cite {harneyb}, but a useful
approximation.

\section{Analysis of composite ensembles}
\label{anal} 

In the present Section we address the unfolding procedure for a large
number of short sequences of level spacings. We do so for two cases:
We study short sequences of levels that do not (do) possess the
rigidity of GOE spectra. This is done in Subsections~\ref{surr}
(\ref{4.2}, respectively). We show that unfolding leads to an
overestimation of the chaoticity parameter in the first but not in the
second case. Thus, long--range order is important for a reliable
unfolding. In Subsection~(\ref{expl}) we comment on this point.

\subsection{The surrogate ensemble}
\label{surr}

Short sequences that lack GOE rigidity are taken from what we call the
surrogate ensemble, a set of statistically independent spacings
obtained with the help of a random--number generator. With this
ensemble we also test (in Section~\ref{test}) the method of statistical
analysis defined in Section~\ref{baye}.

\subsubsection{The uniform surrogate ensemble}
\label{unifsurr}

To obtain the uniform surrogate ensembles, we generate a set of
random numbers $r_j,\, j=1\dots l,$ from a distribution with unit
density between $0$ and $1$. We equate the cumulative spacing
distribution $W$ of Eq.~(\ref{4}) (with $q$ and $Q$ given by
Eqs.~(\ref{7}) and (\ref{11}), respectively) with each of these random
numbers. The associated spacing $s_j$ is then obtained by solving
Eq.~(\ref{4}) for $s$. Setting the  chaoticity parameter to $f_t =
0.60$, one obtains 
\begin{equation}
s_i=-0.482+\sqrt{0.233-2.41\ln (1-r_i)}\, .
\label{24}
\end{equation}
In this way we generate three surrogate ensembles of 50, 100, and
200 spacings, respectively.

In each ensemble these spacings are statistically independent, have
mean spacing unity, and follow the distribution Eq.~(\ref{10}). The
expression ``uniform  ensemble'' refers to the uniform statistical
properties of the spacings. A spectrum constructed from a sequence of
such spacings lacks the rigidity of GOE spectra.

The value $f_t$ is referred to as the ``true'' value of $f$. It is
checked in Section~\ref{test} below whether a statistical analysis of
short sequences of spacings does reproduce this value.

\subsubsection{Comparison of two statistical inferences}
\label{test} 

We use the three surrogate ensembles to compare two methods of
statistical inference: (i) We fit the conditional
distribution~(\ref{10}) to a histogram of the spacings and (ii) we use
the Bayesian method.

(i) The bin size for the histograms was taken once equal to 0.2 and
once equal to 0.3. We determine the chaoticity parameter $f$ for each
of the six histograms by a chi--squared fit to the NNS
distribution~(\ref{10}). The results are given in Figs. 1 and 2
and in Tab. \ref{tab:1}.
\begin{table}
\begin{tabular}{|l|l|l|l|l|}
\hline
Spectrum & Analysis method & $N=50$ & $N=100$ & $N=200$ \\ \hline
Uniform surrogate ensemble & 0.2-bin histogram & $0.90\pm 0.12$ &
$0.75\pm 0.11$ & $0.71\pm 0.06$ \\ \hline
& 0.3-bin histogram & $0.90\pm 0.14$ & $0.80\pm 0.11$ & $0.72\pm 0.06$
\\
\hline
& Bayesian & $0.67\pm 0.14$ & $0.61\pm 0.13$ & $0.63\pm 0.07$ \\
\hline
\end{tabular}
\caption{Bayesian method vs. chi-squared fit}
                                          \label{tab:1}
\end{table}
\begin{figure}[htbp]
\centering
\includegraphics[width=14cm]{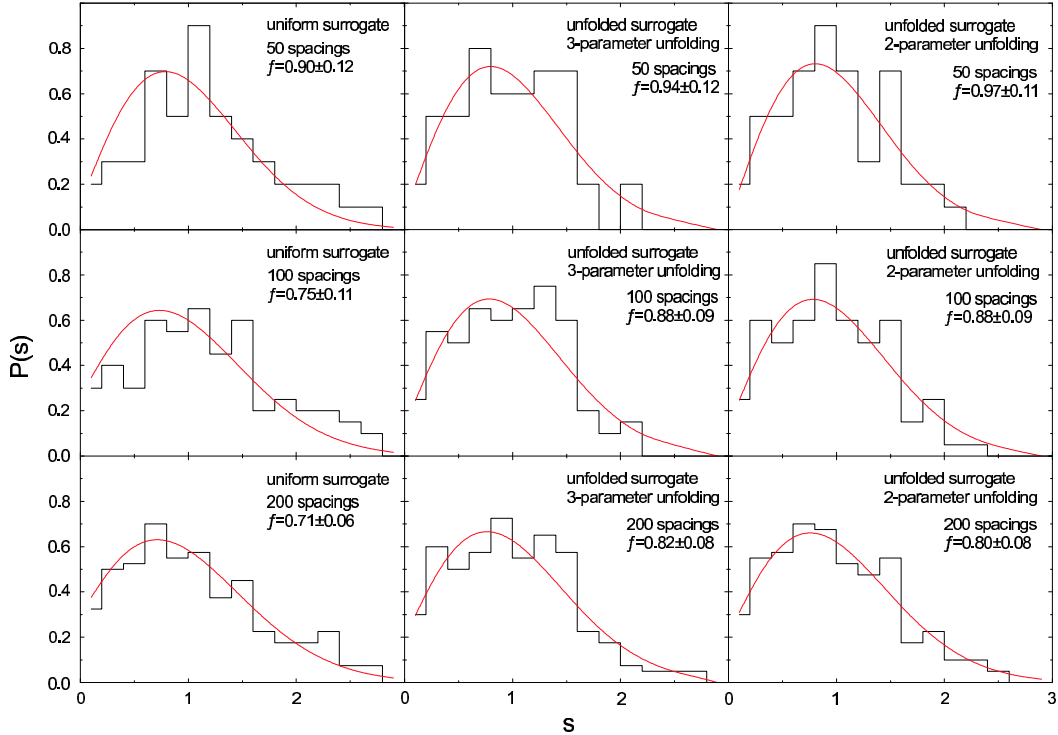}
\caption{Bayesian inference compared to chi-squared fits of the 
distribution \ (3) to the histograms of the
surrogate ensemble. The first column of graphs presents uniform
surrogate ensembles. The second and third columns present unfolded
surrogate ensembles. The bin size is $\Delta s=0.2$.}
 \label{fig:1} 
\end{figure}
\begin{figure}[htbp]
\centering
\includegraphics[width=14cm]{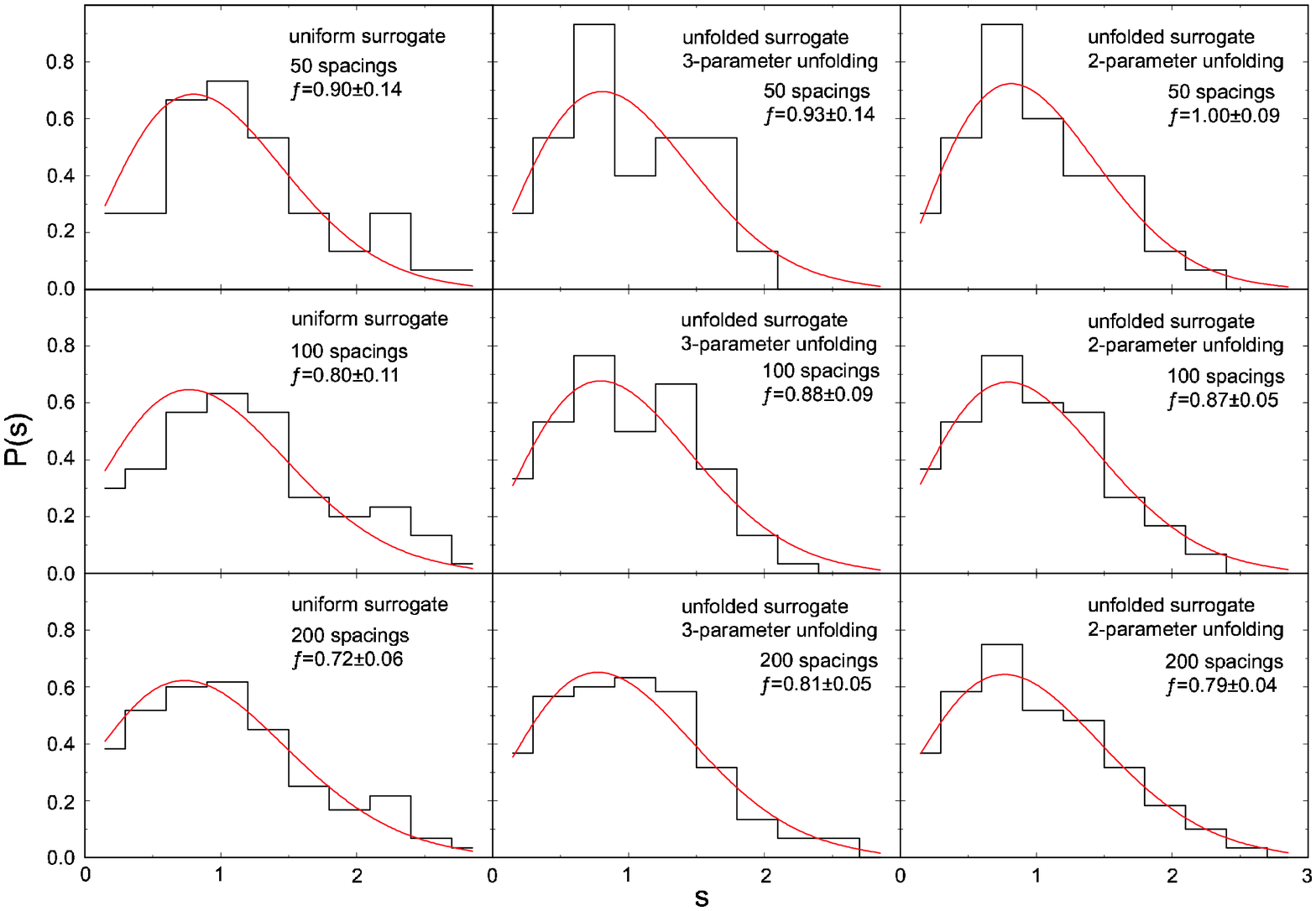}
\caption{Fig. 2. The same as on Fig. 1 except for the bin size $\Delta s=0.3$.}
 \label{fig:2} 
\end{figure}

The Figures show that the shapes of the histograms do not exactly
follow the distribution~(\ref{10}), especially for the smallest
ensemble. Increasing the bin size of the histograms leads to shapes
that better agree with the analytical distribution. One sees this by
comparing Figs.~1 and 2. The chi--squared fit yields values for $f$
that differ from the true value $0.60$, and the difference is
statistically significant, particularly for the large ensemble. This
observation suggests that one should not analyze a small sample of
data by a fit to a histogram. The reason is discussed in Chap.~16 of
Ref.~\cite{harneyb}: The chi--squared fit to a histogram is an
approximation to the Bayesian procedure requiring that the number of
counts in each bin is large.

(ii) The Bayesian method does not require any binning of the data. The
posterior distribution~(\ref{13}) is written in terms of the prior
$\mu (f)$ and a function $\phi (f)$ as
\begin{equation}
P(f|{\bf s})\propto \mu (f)\exp (-N\phi (f))
\label{9}
\end{equation}
for the three uniform surrogate ensembles. We take the prior $\mu$
from the approximation (\ref{16}) in Appendix \ref{app2}. The exponent
$\phi (f)$ is approximated by the cubic form in Eq. (\ref{20}). The 
quality of this approximation is demonstrated in Sec. \ref{4.2.2}.
In the three cases, we determine $\overline{f}$ and the variance 
$\sigma^2$ of Sec. \ref{best}. The results are given in 
Tab. \ref{tab:1} in the
form of $\overline{f}\pm\sigma$. We find that the Bayesian method
correctly reproduces the true value of $f_t=0.60$ within the
statistical error --- for all three distributions involving 50, 100,
and 200 spacings. This indicates that the value of $\overline{f}$ is a
reliable estimate of the true value $f_t$. Tab. \ref{tab:1} contains the cases
where the discrepancy is especially large. However, the very
appearance of a discrepancy and the fact that the chi--squared fit
-- when applied to Poisson statistics -- is less well founded, show
that for small samples of level spacings the Bayesian method is
superior to the fitting of a histogram. Therefore, in the present
article we henceforth report only results obtained with the Bayesian
method (although we have also obtained in parallel the chi--squared
fits).

\subsubsection{Impact of the unfolding procedure}
\label{4.1.3}

Experimental spectra do not possess constant average level density.
Rather the average level density increases with increasing excitation
energy. ``Unfolding'' is the procedure that transforms such a spectrum
into one with constant average level density. In the case of a single
long spectrum unfolding is a standard procedure: It consists in
fitting a slowly varying function $\epsilon (E,\alpha)$ to the
experimental staircase function $N(E)$ of the integrated level density
(a function of energy $E$). Examples of functions $\epsilon$ are given
in Eqs.~(\ref{26},\ref{26a}) below. The fit is obtained by optimizing
a set of parameters $\alpha.$ The function $\epsilon$ depends
monotonically on $E.$ Therefore, one can transform $E$ to $\epsilon.$
With respect to the new energy variable $\epsilon\, ,$ the level
density is equal to unity.

Here we consider an ensemble of spacings given by many short sequences.
For such a ''composite ensemble'', unfolding is not a standard
procedure. One may argue that for composite ensembles, unfolding is
altogether irrelevant because the mean level density changes slowly.
However, sequences of levels of the same spin and parity taken from
the nuclear ground--state domain are examples to the contrary.

To test the impact of the unfolding procedure, we generate short
sequences of levels from the three uniform surrogate ensembles by
arranging the spacings in some order. Each such sequence is then 
artificially folded with a monotonically increasing function of energy.
An unfolding procedure is subsequently applied to each sequence. The
unfolding procedure does not trivially reproduce the initial sequences
and yields the ''unfolded surrogate'' ensembles. Finally, the
chaoticity parameter $f$ is determined.

From the surrogate ensemble containing 50 spacings, we constructed 
8 level sequences. There are 3 sequences of 5 levels, 2 sequences of 6
levels, and one sequence each of 9, of 10 and of 12 levels. The
ensemble containing 100 spacings was arranged into 6 sequences of 5
levels, 4 sequences of 6 levels, two sequences of 7 and two sequences
of 10 levels, and one sequence each of 8, 9, and 12 levels. From the
ensemble containing 200 spacings, we constructed 9 sequences of 5
levels, 6 sequences of 6 levels, 5 sequences of 7 levels, 3 sequences
of 8 levels, 4 sequences of 10 levels, and one sequence each of 9, 17,
and 24 levels. Our choices mirror typical sets of empirical data in
nuclei. We interpret the spacing sequences as level sequences and
identify in each sequence the ground state $E_0$ and the excited
states $E_i$. We then construct ''intermediate'' level sequences $\{
E'_{i}\}$ by the transformation 
\begin{equation}
E'_{i} = aE_{i}^b \, i = 0, \dots, i_{\rm max}
\label{25}
\end{equation}
where $i_{\rm max}$ denotes the number of spacings in the sequence,
and where $a$ and $b$ are positive constants. We refer to the
transition from the energy scale $E$ to the scale $E'$ as to the
``folding'' of the spectrum. 

To make the test as realistic as possible, we determined the 
parameters $a$ and $b$ by fitting the excitation energies of 
the 2$^{+}$ levels of thirty nuclei to Eq.~(\ref{25}). We averaged 
the power $b$ over the thirty cases. We then determined the values
of $a$ that reproduced the positions of the highest 2$^{+}$ level
under consideration in each nucleus, and took the average of $a$. This
procedure yielded $a = 0.36$ and $b = 2.82$. These values were used to
generate the folded spectra.

The intermediate sequences obtained in this manner are unfolded with
the constant--temperature formula~\cite{egidy} 
\begin{equation}
\epsilon (E')=\epsilon_0+\exp\left(\frac{E'-E_0}{T}\right)
\label{26}
\end{equation}
for the cumulative level density. (In the case of experimentally given 
sequences, the energy dependence of the mean level density is not
known a priori. To simulate this situation we chose that dependence to
be different in Eqs.~(\ref{25}) and (\ref{26})). The unfolding was done
in two versions. In a three--parameter version, one searches for the
parameters $\alpha =(\epsilon_0,E_0,T)$ in Eq.~(\ref{26}); in a
two--parameter version, one uses the simpler expression
\begin{equation}
\epsilon (E')=N_1\exp\left(\frac{E'}{T_1}\right)  \label{26a}
\end{equation}
and searches for $\alpha = (N_1,T_1)$. Combining the results in each
case, we obtained the three ``unfolded surrogate'' ensembles of $50,
100,$ and $200$ spacings, respectively, and applied Bayesian
statistics.

The results of this procedure are given in Tab. \ref{tab:2}. We see that the
inferred chaoticity parameter is significantly larger than the true
value $f_t = 0.60$. The ensemble containing 50 spacings appears to be
almost chaotic after unfolding! We explain this as follows. 
\begin{table}
\begin{tabular}{|l|l|l|l|l|}
\hline
Spectrum & Analysis method & $N=50$ & $N=100$ & $N=200$ \\ \hline
Uniform surrogate ensemble & Bayesian & $0.67\pm 0.14$ & $0.61\pm
0.13$ &
$0.63\pm 0.07$ \\ \hline
2-parameter unfolding & Bayesian & $0.88\pm 0.10$ & $0.83\pm 0.08$ &
$0.77\pm 0.06$ \\ \hline
3-parameter unfolding & Bayesian & $0.91\pm 0.08$ & $0.86\pm 0.08$ &
$0.78\pm 0.06$ \\ \hline
\end{tabular}
                                         \label{tab:2}
\caption{Impact of the unfolding procedure on the surrogate ensembles}
\end{table}

Unfolding small sequences yields spacings closer to the mean value
(unity) than would be the case for larger sequences. This is seen by
considering the extreme case of sequences where the number of levels
in each sequence equals the number of parameters $\alpha$. Here,
unfolding would yield a picked--fence spectrum without any
fluctuations of the NNS. Thus, the number of spacings in each sequence
must obviously be larger than the number of parameters used in
unfolding. Still, for small sequences unfolding introduces a bias away
from Poisson and towards the GOE, i.e., the value of $f$ is found
larger than the true one.

To show that our argument applies, we created another unfolded
surrogate ensemble with 50 spacings from only 3 sequences. One of
these sequences is composed of 12 levels, one of 17 levels, and one of
24 levels. The two--parameter unfolding yielded $\overline{f}=0.74\pm
0.13$. This is in better agreement with the true value than the result
$\overline{f} =0.88\pm 0.10$ from the 14--sequence ensemble given in 
Tab. \ref{tab:2}. Correspondingly, the three--parameter unfolding yielded 
$\overline{f}=0.73\pm 0.13$ instead of $\overline{f}=0.91\pm 0.08$ in 
Tab. \ref{tab:2}. 

The values of $\overline{f}$ obtained in Tab. \ref{tab:2} from the unfolded 
surrogate ensembles decrease as the number of spacings increases although 
they remain different from the true value of $f_t=0.60$, even in the
case of $200$ spacings.

\subsection{The GOE ensemble}
\label{4.2}

We now introduce a composite ensemble of  short sequences of levels
following GOE statistics and investigate the effect of unfolding for
this ensemble. GOE spectra differ from those obtained from the surrogate
ensembles by their spectral rigidity. We shall see that the rigidity
is important for unfolding.

When one adds $N$ independent random spacings (as is done in the case
of the surrogate ensembles), the expectation value $\overline{L}$ of
the length of the spectrum is $\overline{L} = N D$, where $D$ is the
mean spacing, while the variance of $L$ behaves as ${\rm var}(L) =
\overline{L^2} - {\overline{L}}^2 \propto N$, so that
\begin{equation}
\sqrt{{\rm var}(L)}/\overline{L}\propto N^{-1/2} \, .
\label{varL1}
\end{equation}
In contradistinction, when one takes $N$ consecutive levels out of a
GOE spectrum, the relative r.m.s. deviation of the of the length $L$
of that sequence is 
\begin{equation}
\sqrt{{\rm var}(L)}/\overline{L}\propto (\ln N)^{1/2}/ N\, .
\label{varL2}
\end{equation}
This makes a noticeable difference particularly for small $N$.

\subsubsection{The uniform composite GOE ensemble}
\label{4.2.1}

The uniform composite GOE ensemble is constructed by using the results
of the experiment by Alt {\it et al.}~\cite{alt}. These authors
measured the resonance spectra of a pair of electromagnetically
coupled superconducting microwave resonators each having the shape of
a quarter of a Bunimovich stadium billiard. The complete spectra of
the two stadia contain 608 and 883 resonances, respectively. We
consider the case where the resonators are uncoupled. Then the
measured spectrum is a superposition of two independent GOE spectra.
It has been unfolded as a whole by Alt {\it et al.}~\cite{alt}. The
resulting spectrum was uniform and the spacing distribution was well
described by Eq.~(\ref{10}) with the true value $f_t = 0.52$, see
Ref.~\cite{dembo}. We ignored the low--energy part of the complete
spectrum and used levels No. 59--463 to generate three ``uniform GOE
ensembles'' of $50, 100$, and $200$ spacings, respectively. The
composition of these ensembles is exactly the same as in
Section~\ref{4.1.3}: The $50$--spacing ensemble, for instance, consists
of 8 sequences of (consecutive) levels with 3 sequences of 5 levels,
2 sequences of 6 levels, one sequence of 9, 10, 12 levels,
respectively, etc. Care was taken to separate each sequence from the
next by a gap of at least 5 levels in the original spectrum of
Ref.~\cite{alt}. This was done in order to reduce or even remove the
correlations between levels belonging to different sequences: The
long--range order of GOE spectra does not imply strong correlations
between spacings that are separated by a few levels in the original
spectrum.

The construction just described corresponds to case (iii) of
Section~\ref{nns}: The sequences are taken from a chaotic quantum
system with an ignored symmetry. We refer to the result as to the 
 ``uniform composite GOE ensemble'' because the sequences contained in
it uniformly possess GOE statistics with average level spacing unity.
We first convince ourselves that $f$ can be correctly inferred from
this ensemble (Section~\ref{4.2.2}). In a second step, the sequences
are folded and again unfolded. This produces the ``unfolded composite
GOE ensemble''. We expect that this ensemble represents typical
experimental situations. We determine the influence of the unfolding
procedure on $f$ by comparing the result with the true value $f_t$.

\subsubsection{Analysis of the uniform composite GOE ensemble}
\label{4.2.2}

To determine the chaoticity parameter $f$ for the three uniform 
composite GOE ensembles, we have expressed the posterior
distribution~(\ref{13}) by the function $\phi (f)$ via Eq.~(\ref{9})
as in Section~\ref{test}. Again $\phi (f)$ was approximated by the
cubic form Eq.~(\ref{20}). The quality of this approximation is
demonstrated in Fig.~3. A similar picture holds for the cases treated
in Section~\ref{test}. We determined $\overline{f}$ and the variance
$\sigma^2$, see Section~\ref{best}. The results are given in Tab. \ref{tab:3}
under the label ``uniform GOE ensemble'' in the form $\overline{f} \pm
\sigma$ . We find the true value $f_t=0.52$ for all three uniform GOE
ensembles involving 50, 100, and 200 spacings.
\begin{table}
\begin{tabular}{|l|l|l|l|l|}
\hline
Spectrum & Analysis method & $N=50$ & $N=100$ & $N=200$ \\ \hline
Uniform GOE  & Bayesian & $0.56\pm 0.14$ & $0.52\pm 0.10$ & $0.55\pm 0.07$
\\ \hline
2-parameter unfolding & Bayesian & $0.59\pm 0.14$ & $0.56\pm 0.09$ &
$0.58\pm 0.09$ \\ \hline
3-parameter unfolding & Bayesian & $0.62\pm 0.13$ & $0.58\pm 0.09$ &
$0.62\pm 0.06$ \\ \hline
\end{tabular}
\caption{Impact of the unfolding procedure on the GOE ensembles}
                                         \label{tab:3}
\end{table}
\begin{figure}[htbp]
\centering
\rotatebox{270}{
                \includegraphics[width=10cm]{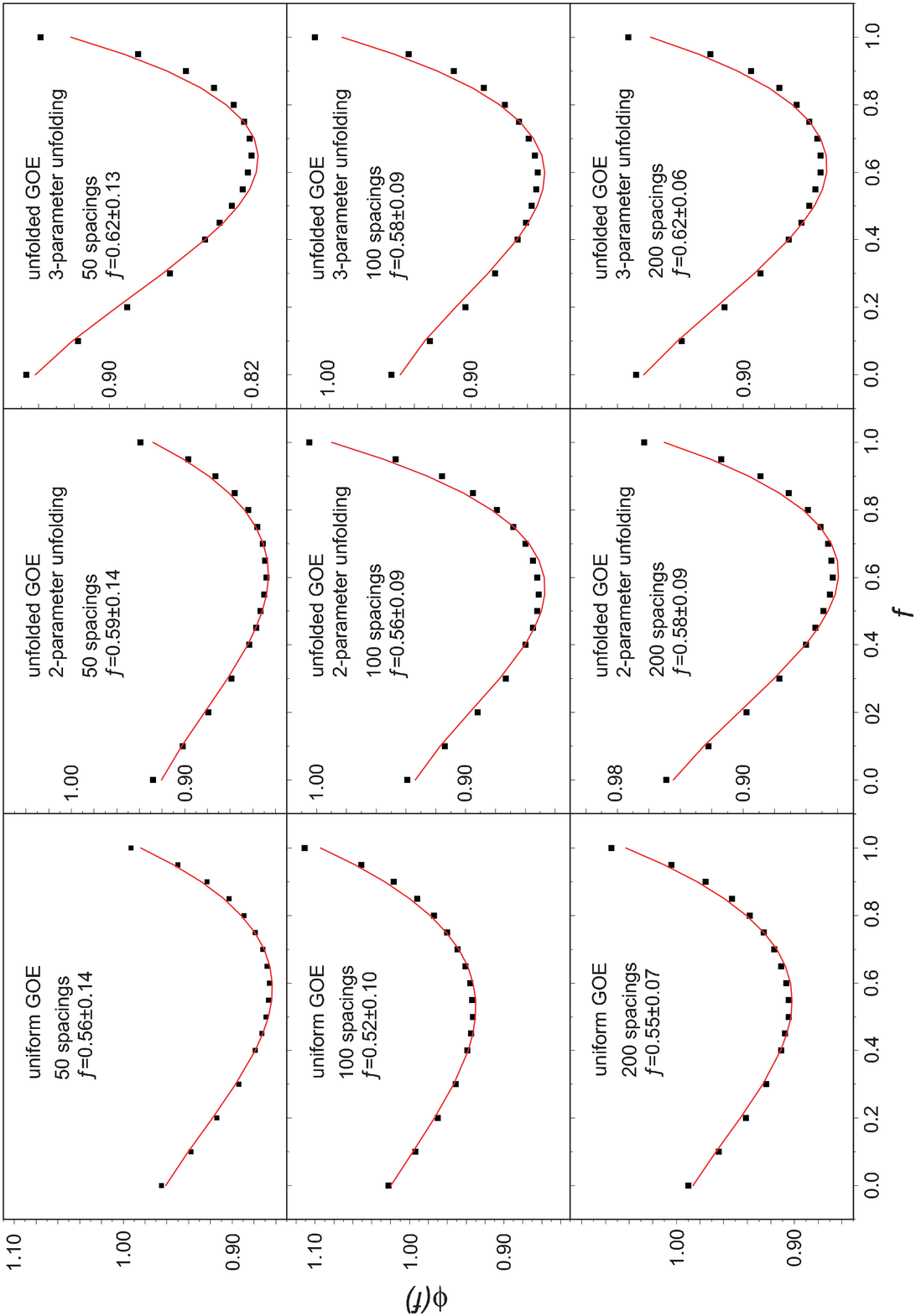}
               }
\caption{The posterior distribution (10) in the approximation (\ref{20}).}
 \label{fig:3} 
\end{figure}

GOE spacings are correlated, especially those that closely follow each
other in a given sequence. Using Eq.~(\ref{12}), we have neglected
these correlations. Our results indicate that such correlations are 
immaterial for the determination of $f$.

\subsubsection{Impact of the unfolding procedure}
\label{4.2.3}

We determine the impact of unfolding the sequences in the GOE ensemble. 
Let $\{E_i\}$ be a sequence of levels taken from the uniform composite
GOE ensemble. We construct an intermediate sequence $\{E'_i\}$ by 
folding $\{E_i\}$ with the function of Eq.~(\ref{25}). The parameters
$a$ and $b$ are as in Section~\ref{4.1.3}. The intermediate sequence is
unfolded with the help of formula~(\ref{26}) (for 3--parameter
unfolding) or (\ref{26a}) (for 2--parameter unfolding). This yields an
unfolded sequence $\{\epsilon_i\}$. The ``unfolded composite GOE
ensemble'' is composed of all unfolded sequences. From the unfolded
composite GOE ensembles, the chaoticity parameter $f$ is inferred.
The results are given in Tab. \ref{tab:3}. We conclude the following:

(i) Inferring $f$ after 2--parameter or 3--parameter unfolding produces 
statistically indistinguishable results. This confirms the
insensitivity of the final ensemble of spacings to the form of the
unfolding function, a hypothesis which has always been implicitly
made in previous analyses. It is indeed a condition for the validity
of the statistical analysis of spectral fluctuations. We do not
investigate this point further because the insensitivity is known from
previous studies.

(ii) In the unfolded GOE ensembles, the true value of $f$ is correctly
inferred (within the error given by the analysis). It seems that the
rigidity of the spectrum of the uniform GOE ensemble guarantees that
the unfolding procedure only weakly alters the fluctuation properties
of the NNS. In the next subsection, we offer an explanation for this
somewhat surprising fact.

\subsection{A better way to infer $f$}
\label{expl}

In our analysis, we have separated the unfolding procedure (which was
done first) from the procedure of inferring $f$. However, unfolding a
sequence of levels amounts to inferring the parameters $\alpha$ of
Eqs.~(\ref{26}) and (\ref{26a}) by way of a fit. The fit does not
determine the value of $\alpha$ precisely. Rather, it is itself a
procedure of statistical inference and --- speaking in terms of 
Bayesian statistics --- it produces a distribution of $\alpha$. We
have disregarded this fact. Instead we have used the best--fit--value
of $\alpha$ in the subsequent estimate of $f$.

A more satisfactory procedure would be the following: One starts from
the distribution $p({\bf s}|f,{\bf\alpha})$ of the spacings $\bf s$
conditioned by the chaoticity parameter $f$ and the parameters $\bf
\alpha$ for all the sequences contained in the composite ensemble
(surrogate or GOE). Hence, $p$ now is the distribution of the raw data
before applying any transformation. The posterior distribution of $f$
is determined after integration over the $\alpha$. In this way one
would not consider $\alpha$ as given; rather one would infer $f$ after
projecting the distribution of all unknown parameters unto the
subspace of the interesting one. In fact, the integration over the
uninteresting parameters is somewhat more complicated because it
touches the question of a possible ``marginalisation
paradox''. Details are discussed in Chap. 12 of \cite{harneyb}.

At any rate, one expects that $f$ is determined with smaller error
when $\bf\alpha$ is given than when $\bf\alpha$ is integrated over.
For the example of  a simple Gaussian model this is demonstrated in
Appendix \ref{app3}.

We have not performed the integration over the uninteresting
parameters $\bf\alpha$. To the best of our knowledge nobody has taken
such a course so far. This is probably because the method would
require a considerable effort --- especially for a composite ensemble
where $\bf\alpha$ is high--dimensional. We conjecture that the
integration over $\bf\alpha$ would yield error intervals for $f$ that
would cover the true value in both cases, the case of the
surrogate ensemble and the case of the GOE ensemble.

This would mean that in the case of the GOE ensemble and under the
simplifying assumption that $\bf\alpha$ is known, we have found 
reasonable error intervals. In the case of the surrogate ensemble,
our simplification seems to fail and we have obtained incorrect (i.e.,
too small) error intervals. Our hypothesis can be made plausible by
comparing two spectra containing the same number of levels. One of the
spectra displays long--range stiffness, the other, local clustering of
levels. The first spectrum allows for a clearer separation between
fluctuations and secular variations than the second one because local
clustering may be confused with a secular variation. The fitting
procedure would react to this difference and in the first case ascribe
a smaller error to $\bf\alpha$ than in the second case.

\section{Summary and Conclusions}
\label{summ} 

In the present paper we have tested the reliability of a statistical
analysis of spectral information contained in many short sequences of
levels. We have used three uniform surrogate ensembles with known
chaoticity parameter $f=f_t$ containing 50, 100 and 200 spacings,
respectively. The surrogate ensembles are due to a random--number
generator, with spacings that are free of the rigidity of chaotic
quantum spectra. The GOE ensembles were constructed from one long
experimental spectrum with chaotic quantum behaviour, unit mean level
spacing, and known chaoticity parameter $f_t$. The spectrum was cut
into short sequences. The sequences were made statistically
independent by disregarding a sufficient number of levels between 
them. We constructed three uniform composite GOE ensembles with 50,
100, and 200 spacings, respectively. (We believe that these ensembles
simulate usual experimental cases). In both cases (the surrogate
ensembles and the GOE ensembles), the method of Bayesian inference
reproduces the true value $f_t$ within the relevant error.  

We have investigated the reliability of the unfolding procedure for
composite ensembles of levels consisting of many short sequences. This
was done by folding sequences of levels constructed from both, the
uniform ensembles and the GOE ensembles, with a monotonically
increasing function of energy. The resulting sequences were then
unfolded with the help of a different unfolding function. The NNS
distributions obtained in this way were combined to form unfolded
ensembles. The resulting chaoticity parameters were determined and
compared to the known true value $f_t$.

The chaoticity parameter $f$ inferred from the unfolded surrogate 
ensembles is too large by about 15\% for the ensembles containing 50 
and 100 spacings, and by about 7\% for the one containing 200 spacings.
These discrepancies are statistically significant. We have shown that
the overestimate is due to the occurrence of many sequences containing
a small number of levels each and would be alleviated if a few long
sequences were available instead. Generically, however, unfolded
surrogate ensembles provide a bias towards full chaoticity. This is
not the case for the GOE ensembles where the true value $f_t$ is
correctly inferred within the statistical error.

In Section~\ref{expl}, we have suggested an explanation for the fact
that unfolded sequences of levels without long--range rigidity
significantly overestimate the chaoticity parameter $f$ while unfolded
sequences of levels that do obey the long--range order of GOE spectra
yield correct estimates of $f$.

To obtain the present results, we have used the one--parameter
conditional distribution of Section~\ref{prop} in order to interpolate
between regular and chaotic systems. Other models of the NNS
distribution --- which we could have used as well --- are expected to
lead to similar conclusions because (i) the Bayesian method is by
definition free from binning the observable and (ii) the explanation
in Section~\ref{expl} is independent of the specific conditional
distribution of Section~\ref{prop}.

Our results provide a justification for previous analyses of atomic,
molecular, and nuclear level statistics where similar composite
spectra were considered. We have in mind the analyses of NNS
distributions of experimental nuclear spectra in the ground--state
domain carried out, for instance, in Refs.~\cite
{aw,egidy,mitchell,raman,shriner,as,garrett,enders,shriner1,abu02},
and the analyses of composite spectra in atoms~\cite 
{rosenzweig,camarda,flambaum} and
molecules~\cite{haller,zimmermann,leitner}. 
Most of these analyses in nuclear systems involve composite ensembles
that combine levels from different nuclei. Even in the few cases where
the levels under investigation are taken from a single nucleus, levels
of different spin or parity are combined to obtain reliable statistics.

\bigskip Acknowledgement \newline
The authors thank Professor J. H\"{u}fner for useful discussions. 
They thank Prof. A. Richter and Dr. C. Dembowski for communicating the
data used to generate the GOE ensembles. A.Y.A.-M.
and M.H.S. acknowledge the financial support granted by Internationales
B\"{u}ro, Forschungszentrum J\"{u}lich, that permitted their stay at the
Max--Planck--Institut f\"{u}r Kernphysik, Heidelberg.

\newpage
\appendix
\section{Level spacing distribution for composite spectra}
\label{app1}

Rosenzweig and Porter~\cite{rosenzweig} considered a spectrum $S$
which can be represented as a superposition of $m$ independent 
sub--spectra $S_{j}$ each having fractional level density $f_{j}$, 
with $j = 1\dots m$, and with $0<f_{j}\leq 1$ and $\sum_{j=1}^{m} f_j
= 1$. Let $p_{j}(s)$ denote the NNS distribution for the sub--spectrum
$S_{j}$ with $j=1\dots m$ and $p(s)$ the NNS distribution of the
spectrum $S$. We define the associated gap functions 
\begin{equation}
E_{j}\left(s\right) =\int_{s}^{\infty }ds^{\prime}
\int_{s^{\prime}}^{\infty}p_{j}(x)dx
\label{gapj}
\end{equation}
for the sub--spectra and 
\begin{equation}
E\left(s\right) =\int_{s}^{\infty }ds^{\prime } \int_{s^{\prime
    }}^{\infty} p(x)dx
\label{gap}
\end{equation}
for the spectrum $S$. Mehta \cite{mehta} has shown that
\begin{equation}
E\left(s\right) =\prod_{j=1}^{m} E_{j}\left( f_{j}s\right)\, .
\label{1}
\end{equation}
Given $E\left(s\right)\, ,$ one can find $p(s)$ by taking the second
derivative of $E\left(s\right)\, . $ The derivation of the exact
expression for $p(s)$ can be found in Appendix A.2, p. 402, of
Ref.~\cite{mehta}. We aim at an approximate evaluation of
$E\left(s\right)$ and, thence, of $p(s)\, .$

We write the cumulative spacing distribution 
\begin{equation}
W(s)=1-\frac{dE(s)}{ds}
\label{cumu}
\end{equation}
in the form 
\begin{equation}
W(s)=1+\exp\left\{\theta\left(s\right) +\ln\left[-d\theta (s)/ds
\right] \right\}\, ,
\label{3}
\end{equation}
where 
\begin{equation}
\theta\left(s\right) =\sum_{j=1}^{m} \ln\left[E_{j}\left(f_{j}s\right)
\right]\, .
\label{Theta}
\end{equation}
In Refs.~\cite{as1}, a simple expression for $W(s)$ was obtained by
expanding the exponent in Eq.~(\ref{3}) in powers of $s$\ and
neglecting terms of higher order than the second. This procedure was
motivated by the fact that $p(s)$ is mainly determined by short--range
level correlations. We follow this procedure and obtain 
\begin{equation}
W(s)=1-\exp\left(-qs-\frac{\pi}{4}Qs^{2} \right)\, ,
\label{4}
\end{equation}
where 
\begin{equation}
q=1-\sum_{j=1}^{m} f_{j}^{2}\left[1-p_{j}\left(0\right) \right]
\label{5}
\end{equation}
and 
\begin{eqnarray}
Q &=&\frac{2}{\pi } \biggl(\sum_{j=1}^{m} f_{j}^{2}\left[1-p_{j}(0)
\right] +\left\{\sum_{j=1}^{m} f_{j}^{2}\left[ 1-p_{j}(0) \right]
\right\} ^{2}
\nonumber \\
& & \quad\quad -\sum_{j=1}^{m}
f_{j}^{3}\left[2-3p_{j}(0)-p_{j}^{\prime }(0) \right] \biggr)\, .
\label{6}
\end{eqnarray}
The Wigner surmise for each of the $p_{j}$'s implies $p_{j}(0)=0$ and
$p_{j}^{\prime }(0)=\pi /2$. The parameter $q$ --- given by
Eq.~(\ref{5}) --- becomes 
\begin{equation}
q=1-f\, ,
\label{7}
\end{equation}
where 
\begin{equation}
f=\sum_{j=1}^{m} f_{j}^{2}\, .
\label{8}
\end{equation}
Since $\sum_{j} f_{j}=1\, ,$ we cannot use the mean value of the
$f_{j}$'s as a measure of the mean fractional density. The parameter
$f$ with $0<f\leq 1$ defined in (\ref{8}) is the next--best choice. We
refer to $f$ as to the mean fractional level density for the
superimposed sequences. The parameter $Q$ is, in principle, given by
Eq.~(\ref{6}). We will, however, replace that expression by another
one which we construct as follows.

The NNS distribution is obtained by differentiating the approximate
expression (\ref{4}) with respect to $s$, 
\begin{equation}
p(s)=\frac{{\rm d}W(s)}{{\rm d}s}\, .
\label{31}
\end{equation}
In going from Eq.~(\ref{3}) to Eq.~(\ref{4}), we have neglected higher
powers of $s$ in the expansion of $\ln [1-W(s)]$. This neglect
entails, however, that the distribution~(\ref{4}) does not satisfy the
condition of unit mean spacing
\begin{equation}
\int_{0}^{\infty } xp(x)dx = \int_{0}^{\infty }\left[1-W(x)\right]dx
= 1 \, .
\label{9a}
\end{equation}
In order to satisfy this condition we determine the parameter $Q$ from
Eq.~(\ref{9a}) while keeping Eq.~(\ref{7}) for the parameter $q$. We
do so in order to maintain the correct behavior of a collection of
independent GOE subsequences at small values of $s$. Hopefully, this
approximation will take into account some of the effects of the
neglected terms in the power--series expansion of the logarithm. The
proposed NNS distribution of the composite spectrum is then given by 
\begin{equation}
p(s,f)=\left[1-f+Q(f)\frac{\pi s}{2} \right]
\exp\left[-\left(1-f\right)s -Q(f)\frac{\pi s^{2}}{4} \right] \, ,
\label{10a}
\end{equation}
where $Q(f)$ is defined by the condition~(\ref{9a}). This procedure
yields an implicit relation between $Q$ and $f$ which involves a
complementary error function. We have numerically solved the implicit
equation and obtained $Q(f)$ for $f$ in the interval of $0.1 \leq f
\leq 0.9 \, .$ The resulting solution was approximated by the
parabolic relation
\begin{equation}
Q(f)=f\left( 0.7+0.3f\right)\, .
\label{11}
\end{equation}
With this approximation, the mean spacing differs from unity by less
than 0.5\%. The distribution~(3) coincides with the exact expression
up to the 6th decimal digit. The exact values were obtained by doubly
differentiating Eq.~(\ref{1}), see Ref.~\cite{mehta}.

\section{Details on the posterior distribution}
\label{app2}
 
The prior distribution Eq.~(\ref{15}) was evaluated numerically after
inserting Eq.~(\ref{12}) into Eq.`(\ref{15}). The result was
approximated by the sixth--order polynomial 
\begin{eqnarray}
\mu \left( f\right) &=& 1.975-10.07f+48.96f^2-135.6f^3  \nonumber \\
& &+205.6f^4-158.6f^5+48.63f^6 \ .
\label{16}
\end{eqnarray}

The distribution $p\left( {\bf s}\left| f\right. \right) $\ assumes
very small values even for only moderately large values of
$N$. Therefore, the accurate calculation of the posterior distribution
requires some care. In order to simplify the calculation, we have
rewritten Eq.~(\ref{12}) in the form
\begin{equation}
p\left( {\bf s}\left| f\right. \right) =e^{-N\phi (f)}\, ,
\label{17}
\end{equation}
where 
\begin{eqnarray}
\phi (f)&=& (1-f)\langle s\rangle +\frac{\pi
  }{4}f\left(0.7+0.3f\right) \langle s^{2}\rangle  \nonumber \\
& &-\langle\ln\left[1-f+\frac{\pi }{2}f\left(0.7+0.3f\right)s \right]
\rangle\, .
\label{18}
\end{eqnarray}
Here, the notation 
\begin{equation}
\langle x\rangle =\frac{1}{N}\ \sum_{i=1}^{N} x_{i}
\label{19}
\end{equation}
has been used. We find that the function $\phi (f)$ has a pronounced
absolute minimum, say at $f=f_{0}$. This minimum provides the maximum
of the posterior distribution $P$ which is of interest in the
neighborhood of $f_0$. There one can represent $P$ by parameterizing
$\phi$ in the form of a third--order polynomial,
\begin{equation}
\phi (f)=A+B\left(f-f_{0}\right)^2+C\left(f-f_0\right)^3\, .
\label{20}
\end{equation}
The parameters $A,B,C$ and $f_{0}$ are implicitly defined by
Eq.~(\ref{18}).

In the analysis of the NNS distributions for the coupled microwave
resonators~\cite{dembo}, the number of spacings for each coupling was 
so large ($N \simeq 1500$) that a Gaussian distribution 
\begin{equation}
P_{{\rm {G}}} \left( f \left| \right. {\bf s} \right) \propto \exp
\{-NB (f - f_{0})^{2} \}
\label{21}
\end{equation}
was found to describe the $f$--dependence of the posterior
$P\left(f\left|{\bf s} \right.\right)$ very well. Indeed, 
for a sufficiently large number $N$ of data the posterior should 
approach a Gaussian. Although this is not a consequence of the
Central Limit Theorem, the proof of this statement is similar.
Therefore, the posterior distributions obtained in that analysis
were Gaussians characterized by a mean value $f_0$  and the 
variance $\sigma_0=1/\sqrt{2NB}$. The present analysis, however,
addresses NNS distributions that involve a considerably smaller
number of spacings. Therefore, we cannot further simplify the
approximation~(\ref{20}) and arrive at 
\begin{equation}
P\left( f \left| {\bf s} \right. \right) 
   = c\,\mu\left(f\right) 
     \exp\left(-N\left[B\left(f-f_0\right)^2+C\left(f-f_0\right)^3 
                 \right] 
         \right),\, 0\le f\le 1\, .
\label{22}
\end{equation}
Here, $c = e^{-NA}\left/M({\bf s})\right.$ is the new normalization
constant.

\section{Integration over an uninteresting parameter}
\label{app3}

We expand on the integration over uninteresting parameters briefly
discussed in Subsection~\ref{expl}. Let the model $p(x|\xi )$ be
conditioned by two parameters,
\begin{equation}
\xi=\left(\begin{array}{c}\xi_1\\
                          \xi_2
          \end{array}
    \right)\, .
\label{A3.1}
\end{equation}
Only $\xi_1$ is interesting. The precision with which one can infer
the value of $\xi_1$ depends on one's knowledge of $\xi_2$. We
distinguish two cases of prior knowledge: (i) $\xi_2$ is known to have
the value $\alpha$ and (ii) $\xi_2$ is unknown. In the first case,
$\xi_2 =\alpha$ is simply inserted into the model so that $\xi_1$ is
inferred from $p(x|\xi_1; \xi_2  =\alpha)$. In the second case,
$\xi_1$ is inferred after $\xi_2$ has been integrated over. The
precision with which $\xi_1$ can be obtained, is better in the first
case than in the second.

As an example we consider the Gaussian model
\begin{equation}
p(x|\xi )\propto\exp\left(-(x-\xi )^T(2C)^{-1}(x-\xi )
                    \right)\, .
\label{A3.2}
\end{equation}
Here, the data $x$ form a 2--dimensional vector
\begin{equation}
x=\left(\begin{array}{c}x_1\\
                        x_2
          \end{array}
  \right)
                        \label{A3.3}
\end{equation}
as do the parameters $\xi$. The 2--dimensional correlation matrix 
$C$ contains the variances $C_{11}, C_{22}$ and the correlation
coefficient $C_{12}$. (This model has also been discussed in
Chap. 12.2.2 of \cite{harneyb}).

(i) When $\xi_2$ is known to have the value $\alpha$, one obtains the
Gaussian model
\begin{equation}
p(x|\xi_1;\xi_2=\alpha )
    \propto\exp\left(-\left((2C)^{-1}\right)_{11}
                \left[(x_1-\xi_1+(x_2-\alpha)C^{-1}_{12}/C^{-1}_{11})^2
                \right]
               \right)
\label{A3.4}
\end{equation}
conditioned by $\xi_1$. The posterior distribution of $\xi_1$ is
Gaussian with variance 
\begin{eqnarray}
\left((C^{-1})_{11}
\right)^{-1} &=& \left({C_{22}\over\det C}\right)^{-1}\nonumber\\
             &=& C_{11}-C_{12}^2/C_{22}.
\label{A3.5}
\end{eqnarray} 

(ii) Integration of Eq.~(\ref{A3.2}) over $\xi_2$ yields the Gaussian
model
\begin{eqnarray}
q(x|\xi_1) &=      & \int {\rm d}\xi_2\, p(x|\xi_1,\xi_2)\nonumber\\
           &\propto&  \exp\left(-(2C_{11})^{-1}(x_1-\xi_1)^2
                          \right)
\label{A3.6}
\end{eqnarray}
condioned by $\xi_1$. The posterior distribution of $\xi_1$ is 
Gaussian with variance $C_{11}$. 

The variance ($C_{11}$) of the second case is larger than the
variance~(\ref{A3.5}) of the first case. The variances agree if the
correlation $C_{12}$ vanishes. Then the problem factorises with
respect to the parameters $\xi_1$ and $\xi_2$.

\end{document}